\documentclass[prb,twocolumn,superscriptaddress, showpacs,floatfix]{revtex4}

\usepackage[dvips]{graphicx}
\usepackage{amssymb, amsmath, bbm,}

\def\be{\begin{equation}}
\def\ee{\end{equation}}
\def\beq{\begin{eqnarray}}
\def\eeq{\end{eqnarray}}

\newcommand{\bs}{\boldsymbol}


\begin{document}
\title{Diamond chains with multiple-spin exchange interactions}
\author{N. B.  Ivanov}
\affiliation{Institute of Solid State Physics,  Bulgarian Academy of
Sciences, Tzarigradsko chaussee 72, 1784 Sofia, Bulgaria}
\author{J. Richter}
\affiliation{Institut f\"ur Theoretische Physik, Universit\"at Magdeburg,
PF 4120, D-39016 Magdeburg, Germany}
\author{J. Schulenburg}
\affiliation{Universit\"{a}tsrechenzentrum,
             Universit\"{a}t Magdeburg, D-39016 Magdeburg, Germany}

\date{\today}
\begin{abstract}
We study the phase diagram of a symmetric spin-$1/2$  Heisenberg 
diamond chain  with additional cyclic four-spin exchange interactions. 
The presented analysis supplemented by numerical exact-diagonalization 
results for  finite periodic clusters implies a rich phase diagram 
containing, apart from standard magnetic and spin-liquid phases, 
two different tetramer-dimer phases  as well as
an exotic four-fold degenerate dimerized phase.   
The characteristics of the established spin phases as well as
the  nature of quantum phase transitions are discussed, as well.  

\end{abstract}
\maketitle

\section{Introduction}
Spin systems with cyclic exchange interactions have been  receiving an 
increasing amount of attention in the past few years. These 
interactions are  known to be responsible for the properties 
of the triangular magnetic system composed of $^3$He atoms absorbed on 
graphite surfaces.\cite{roger} Recently, it was 
demonstrated that a certain amount of four-spin exchange could 
explain the neutron-scattering experiments concerning  high-$T_c$ compounds
such as La$_2$CuO$_4$,\cite{coldea}, La$_6$Ca$_8$Cu$_24$O$_{41}$,\cite{brehmer}, 
and  La$_4$Sr$_{10}$Cu$_{24}$O$_{41}$.\cite{notbohm}   
The latter two are spin-ladder compounds where charge fluctuations such as
cyclic hopping processes modify the Heisenberg spin Hamiltonian by
contributing a four-spin interaction term. In particular, it was found
that these interactions substantially modify  the spin triplon gaps and 
frustrate the formation of bound triplon states.\cite{notbohm}
On the theoretical side, in spite of the numerous numerical studies  
predicting rich phase diagrams,\cite{lauchli} 
a number of important questions, concerning the type of spin orderings 
and quantum criticalities realized by the four-spin exchange, 
remain unsettled.\cite{hikihara}   

\begin{figure}[hbt]
\samepage
\begin{center}
\includegraphics[width=7cm]{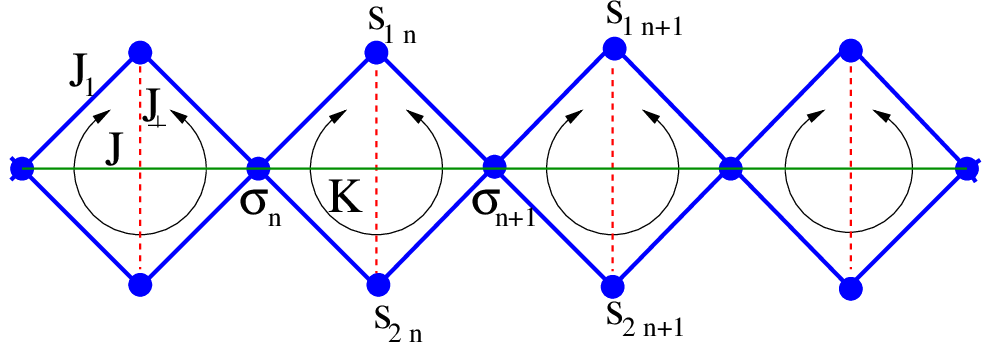}
\caption{\label{dchain} The symmetric diamond chain considered in the paper.
The arrows denote the cyclic four-spin exchange interaction controlled
by the parameter $K$.  
}
\vspace{-.7cm}
\end{center}
\end{figure}

In this paper  we  analyze  the role  of the cyclic
four-spin exchange interactions in the symmetric diamond chain (SDC)
which is one of the simplest spin models admitting multispin
cyclic exchange interactions. The model is defined by the Hamiltonian
(see Fig.~\ref{dchain}) 
\be\label{h}
{\cal H}\!=\!\sum_{n=1}^L h_{n,n+1}\, ,
\ee
where
\beq
h_{n,n+1}\!&=&\!J_1{\bf s}_n\cdot \left( 
\bs{\sigma}_n+\bs{\sigma}_{n+1}\right)
+J\bs{\sigma}_n\cdot\bs{\sigma}_{n+1}\nonumber \\
&+&\!J_{\perp}{\bf s_1}_n\cdot{\bf s_2}_{n}
\!+\! K\left[ \left( {\bf s_1}_n\cdot\bs{\sigma}_n\right)\!
\left( {\bf s_2}_n\cdot\bs{\sigma}_{n+1}\right)\right.\nonumber \\
&+&\!\! \left. \left( {\bf s_1}_n\cdot\bs{\sigma}_{n+1}\right)\!
\left( {\bf s_2}_n\cdot\bs{\sigma}_{n}\right)
\!-\!\left( {\bf s_1}_n\cdot{\bf s_2}_{n}\right)
\left( \bs{\sigma}_n\cdot\bs{\sigma}_{n+1}\right)
\right]\, .\nonumber 
\eeq
Here  ${\bf s_1}_n$, ${\bf s_2}_n$,
and $\bs{\sigma}_n $ are spin-$1/2$ operators defined on the sites
of the  $n$-th elementary cell.  In our model, the standard cyclic 
four-spin exchange  interactions\cite{muller} are slightly generalized
by  including its  bilinear terms  in the exchange parameters 
$J$, $J_1$,  and $J_{\perp}$. In what follows the action and the 
energy are measured in the units of $\hbar$ and $J_1$, respectively.

The frustrated diamond chain\cite{takano} as well as its 
various  modifications, such as the  distorted diamond 
chain\cite{okamoto,mikeska} 
and the so-called $AB_2$ ferrimagnetic chain with Ising and Heisenberg
spins\cite{vitoriano},  have  already been discussed in the literature
in relation  to some  quasi-1D magnetic materials.\cite{rule} 
For the  following analysis it is important to notice that
the cyclic exchange interaction does not violate the local 
symmetry of the Hamiltonian 
under the exchange of the  pair of spins  (${\bf s_1}_n,{\bf s_2}_n$)  
for each  diamond in the SDC.
Thus, in the important case of spin-$1/2$ operators ${\bf s_1}_n$ and
${\bf s_2}_n$,  the system is characterized by $L$  local 
good quantum numbers $s_n=0,1$ ($n=1,2,\dots,L$) related to the composite
spins  ${\bf s}_n={\bf s_1}_n+{\bf s_2}_n$: ${\bf s}_n^2=s_n(s_n+1)$.  
Using this local symmetry and the standard commutation relations for
spin operators, the local Hamiltonian $h_{n,n+1}$
can be represented in the compact form
\beq\label{h2}
h_{n,n+1}&=&\epsilon_n
+{\bf s}_n\cdot\left( \bs{\sigma}_n+\bs{\sigma}_{n+1}\right)
\! +\! J_n\bs{\sigma}_n\cdot
\bs{\sigma}_{n+1}\nonumber \\
&+&\frac{K}{2}\{ {\bf s}_n\cdot\bs{\sigma}_n\, ,\, 
{\bf s}_n\cdot\bs{\sigma}_{n+1}\} \, .
\eeq
Here  $\epsilon_n/J_{\perp}=s_n(s_n+1)/2-3/4$ are fixed
numbers ($-3/4$ or $1/4$) for every sector defined as a 
sequence of the local quantum numbers $\left[ s_1,s_2,\dots,s_L\right]$,
$J_n=J +K/4 - s_n(s_n+1)K/2$, and $\{ A,B \}$ is the anticommutator  
of two operators ($A$ and $B$).  

In the parameter space where the ground state  is characterized by
$s_n=0$ ($n=1,2,\ldots ,L$), the second and fourth terms in Eq.~(\ref{h2})
vanish and the   model is equivalent to the 
spin-$1/2$ Heisenberg chain with an exchange parameter $J+K/4$.
On the other hand, if the ground state  belongs to the sector
$s_n=1$ ($n=1,2,\ldots ,L$), Eq. (\ref{h2}) defines the generic model of
an alternating mixed-spin chain with the simplest three-spin exchange 
interaction. Finally, if the ground state is in the sector [$1,0,1,0,\ldots$], 
the system is reduced to some modification of the well-known orthogonal 
dimer chain.\cite{ivanov1} 
\section{Phase diagram in the parameter space ($K$,$J_{\perp}$)}
\begin{figure}[hbt]
\begin{center}
\includegraphics[width=7.5cm]{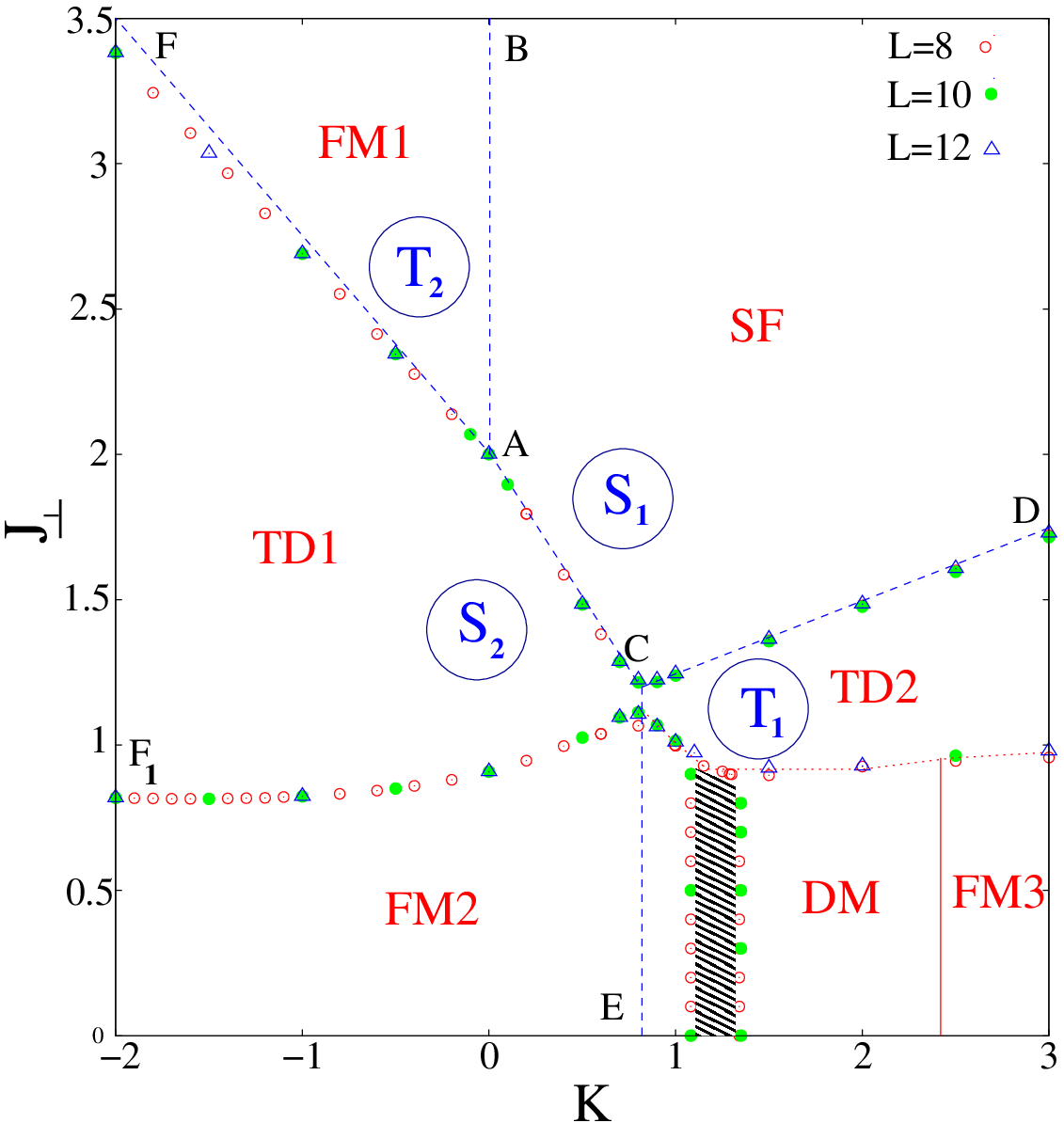}
\vspace{-.4cm}
\caption{\label{ph_diagram} Phase diagram of the model, Eq.~(\ref{h}),  
in the parameter space  ($K,J_{\perp}$) for $J=0$  as  obtained from 
the numerical  diagonalization of periodic clusters containing up 
to $L=12$ unit cells. The dashed lines denote the boundaries
of the single-diamond phase diagram containing the 
states $S_1$, $S_2$, $T_1$, and $T_2$ (see Table 1).  
The following abbreviations for the phases of Eq.~(\ref{h}) are used:
critical spin-fluid state (SF), fully-polarized ferromagnetic state (FM1),
ferrimagnetic state (FM2), tetramer-dimer states (TD1,TD2), dimerized
singlet state (DM), and another ferrimagnetic state (FM3).     
}
\vspace{-.7cm}
\end{center}
\end{figure}

 It is instructive to begin with the
phase diagram of a single diamond composed of the spins 
${\bf s_1}$, ${\bf s_2}$, $\bs{\sigma}_1$, and $\bs{\sigma}_2$.
The diamond eigenstates  (see Table 1) consist of  two singlets ($S_1$ 
and $S_2$), three triplets ($T_1^{\mu}$, $T_2^{\mu}$, and $T_3^{\mu}$;
$\mu = 0,\pm 1$), 
and the quintet $Q^{\mu}$ ($\mu = 0,\pm 1,\pm2$). In Figure~\ref{ph_diagram}, 
the  phase boundaries 
between the  single-diamond ground states ($S_1$, $S_2$, $T_1$, and $T_2$)  
are depicted  by dashed lines.  As discussed below, some of these lines  
coincide  with the  exact phase boundaries of the diamond chain, Eq.~(\ref{h}). 
\begin{table}[h]
\begin{center}
\begin{tabular}{l|lll|l|l}
\hline
      &$s_p$ &\! $s$ & $\sigma$ & \hspace{0.5cm} Eigenvalue 
& \hspace{1cm} Eigenstate\\
\hline
$S_1$ & 0 &\! 0 & 0 & $-\frac{3}{4}(J_{\perp}+J)-\frac{3}{16}K$ &
$t_{\perp}^s t_{||}^s$\\
$S_2$ & 0 &\! 1 & 1 & $-2\! +\! \frac{1}{4}(J_{\perp}\! +\! J)
\!+\!\frac{13}{16}K$ & 
$\frac{1}{\sqrt{3}}(t_{\perp}^+ t_{||}^-
\! +\! t_{\perp}^- t_{||}^+\! -\! t_{\perp}^0 t_{||}^0)$ \\
$T_1^0$ & 1 &\! 1 & 1 & $-1\! +\! \frac{1}{4}(J_{\perp}\! +\! J)\! -
\! \frac{7}{16}K$ &  
$\frac{1}{\sqrt{2}}( t_{\perp}^+t_{||}^-
-t_{\perp}^- t_{||}^+)$ \\
$T_2^{\mu}$ & 1 &\! 0 & 1 & $-\frac{3}{4}J_{\perp}+\frac{J}{4}+\frac{K}{16}$ & 
$t_{\perp}^s t_{||}^{\mu}$, \, $\mu =0,\pm$ \\
$T_3^{\mu}$ & 1 &\! 1 & 0 & $\frac{J_{\perp}}{4}-\frac{3}{4}J+\frac{K}{16}$ & 
$t_{\perp}^{\mu}t_{||}^{s}$, \, $\mu =0,\pm$ \\
$Q^0$   & 2 &\! 1 & 1 &$1\! +\! \frac{1}{4}(J_{\perp}\! +\! J)
\!+\!\frac{K}{16}$ & $\frac{1}{\sqrt{6}}( t_{\perp}^+ t_{||}^-
\! +\! t_{\perp}^- t_{||}^+\! +\! 2 t_{\perp}^0 t_{||}^0)$ \\
\end{tabular}
\caption{\label{table}Eigenvalues and eigenstates  of a single diamond 
composed of the spins ${\bf s}_1$, ${\bf s}_2$, $\bs{\sigma}_1$, 
and $\bs{\sigma}_2$. The eigenstates are  classified according to the 
following good quantum numbers of the  single-diamond cluster:
the total diamond  spin $s_p$, its $z$ component $s_p^z$,  and the diagonal 
spins $s$ and $\sigma$ [${\bf s}={\bf s}_1+{\bf s}_2$ 
and $\bs{\sigma}=\bs{\sigma}_1+\bs{\sigma}_2$: ${\bf s}^2=s(s+1)$,
${\bs\sigma}^2=\sigma (\sigma +1)$]. The symbols
$t^{\mu}$ ($\mu =0,\pm$) stand for the canonical basic states of the 
spin-1 operators, whereas $t^{s}$ denotes  the singlet state of two spin-$1/2$
operators. For brevity, only the  $s_p^z=0$ components of the  triplet 
($T_1^{\mu}$)  and  quintet ($Q^{\mu}$) states are presented.}
\vspace{-.7cm}
\end{center}
\end{table}
\subsection{Phases in the sector 
$[0,0,\cdots ,0]$}
For large enough values of the parameter $J_{\perp}$, 
the off-chain spins  form local dimers, ${\bf s_1}_n\cdot{\bf s_2}_{n}=-3/4$
so that the ground state of the model  belongs to the sector
$[0,0,\cdots ,0]$.  Thus, Eq.~(\ref{h2}) reduces to the form
\be
h_{n,n+1}=-(3/4)J_{\perp}+(J+K/4)\bs{\sigma}_n\cdot\bs{\sigma}_{n+1}. 
\ee
This is the Hamiltonian  of a spin-1/2 Heisenberg chain with the
exchange constant $J+K/4$. The single-diamond line $AB$ (defined by $J+K/4=0$)
coincides with the exact boundary between the fully polarized ferromagnetic 
phase FM1 ( $J+K/4<0$) and the critical spin-fluid  phase SF ( $J+K/4>0$).

Besides the well-documented collective modes, these  phases  exhibit 
additional   single-particle   modes  (related  to the off-diagonal
spins) describing  excited  
(${\bf s_1}_n,{\bf s_2}_n)$ dimers, $s_n=1$.
Being eigenstates of the Hamiltonian, Eq.~(\ref{h}), these 
excitations are completely localized. According to Eq.~(\ref{h2}),  an   
excited dimer (say at $n=L$) is described by the 
Hamiltonian  
\be\label{hi}
{\cal H}_{i}= C+\left(J+\frac{K}{4}\right)
\sum_{n=1}^{L-1}\bs{\sigma}_n\cdot\bs{\sigma}_{n+1}+h_{i}, 
\ee
where $C=-\frac{3}{4}J_{\perp}(L-1)$ and  
$h_{i}$ reads 
\beq\label{hii}
h_{i}&\! =&\! \frac{J_{\perp}}{4}\! +\!\left(J\! -\!\frac{3K}{4}\right)
\bs{\sigma}_L\cdot\bs{\sigma}_{1}\!+\!J_{i}
{\bf s}_L\cdot\left( \bs{\sigma}_L
\! +\!\bs{\sigma}_{1}\right)\nonumber \\
&+&\frac{K}{2}\{ {\bf s}_L\cdot\bs{\sigma}_L\, ,\,
{\bf s}_L\cdot\bs{\sigma}_{1}\},\,\, J_{i}=1.
\eeq
Similar  models,  describing  spin-$S$ impurities in 
spin-1/2 antiferromagnetic Heisenberg chains,   have 
been discussed  in the  literature.\cite{sorensen} In particular, 
the above model may be considered as  a special class of open chains with  
symmetric  couplings of the  end spins ($\bs{\sigma}_{1}$  
and $\bs{\sigma}_{L}$) to the external spin ${\bf s}_L$ ($s_L=1$).

Turning to the region occupied by the SF phase, 
renormalization-group arguments predict that symmetric
perturbations to the open chain are, at most, marginal. 
Thus,  the marginally relevant perturbation 
$J_{i}{\bf s}_L\cdot\left( \bs{\sigma}_L
\! +\!\bs{\sigma}_{1}\right)$ ($J_i>0$) is expected to renormalize 
to $\infty$ in the SF phase. This corresponds to
a fixed point  where the end spins $\bs{\sigma}_{1}$
and $\bs{\sigma}_{L}$ are effectively removed from the chain to screen   
the external  spin ${\bf s}_L$. In terms of the original model, 
Eq.~(\ref{h}), the  spins  $\bs{\sigma}_{1}$, $\bs{\sigma}_{L}$,  
${\bf s_1}_L$, and ${\bf s_2}_L$ form a decoupled  
single-diamond state 
($S_2$)  in the low-energy sector of the spectrum.  
Such local excitations  are relevant  relatively close to the  
phase boundary $AB$,  where 
the coupling $J_{i}{\bf s}_L\cdot\left( \bs{\sigma}_L\! 
+\!\bs{\sigma}_{1}\right)$ dominates the biquadratic term 
in Eq.~(\ref{hii}).

For larger values of the parameter $K$,
the biquadratic exchange  in $h_i$ becomes important. 
Since the energy of the quintet state $Q$  grows up with $K$
(see Table 1), we  concentrate on  the  triplet  state $T_1$ whose  
energy decreases with $K$  and crosses 
the energy level of $S_2$ at $K=4/5$. 
As discussed in Ref.~\onlinecite{sorensen}, 
such a local triplet  state does not correspond to a stable fixed point, 
since the antiferromagnetic interaction of the effective spin-1 impurity 
with the rest of the chain is marginally relevant. Thus,  one  
expects that the couplings to the next two spins in the chain, 
$\bs{\sigma}_{L-1}$ and $\bs{\sigma}_{2}$,  grow up to $\infty$. 
There appears another fixed point where the latter two spins also 
decouple from the chain in order to  screen  the spin of the triplet state 
$T_1$.  Note that in sectors containing more  spin-1 dimers, 
other types of screening are possible, as well.  For example,  the spins of
neighboring  $T_1$ diamonds in  Fig.~\ref{td_state} may be  screened by
forming a  singlet state. As before,
the number of decoupled chain spins is four. We suggest that 
such decoupled singlet states, composed of longer $S_2$ or $T_1$ diamond 
chains, control the observed instability  of the  SF phase upon decreasing
the parameter $J_{\perp}$. Note that similar  complexes of $S_1$ diamonds 
can not produce the instability, since the energy of the resulting product 
state $[\bs{\sigma}_1,\bs{\sigma}_2][\bs{\sigma}_3,\bs{\sigma}_4]\cdots
[\bs{\sigma}_{L-1},\bs{\sigma}_L]$  exceeds the energy of 
the critical phase for arbitrary $J_{\perp}, K>0$.       
\subsection{Tetramer-dimer phases}
\begin{figure}[hbt]
\begin{center}
\includegraphics[width=7.5cm]{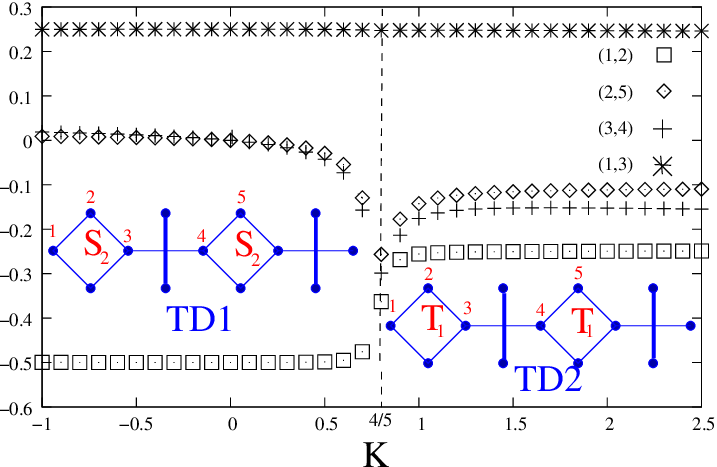}
\vspace{-.4cm}
\caption{\label{td_state} Short-range spin-spin correlations vs. $K$ in the 
tetramer-dimer phases TD1 and TD2  for
periodic chains with $L=8$ cells ($J=0, J_{\perp}=1.15$). 
$(1,2)\equiv \langle \bs{\sigma}_n\cdot\bs{s_1}_n\rangle$,    
$(1,3)\equiv \langle \bs{\sigma}_n\cdot\bs{\sigma}_{n+1}\rangle$,
$(3,4)\equiv \langle \bs{\sigma}_{n+1}\cdot\bs{\sigma}_{n+2}\rangle$,
$(2,5)\equiv \langle\bs{s_1}_n\cdot\bs{s_1}_{n+2} \rangle$.
}
\vspace{-.7cm}
\end{center}
\end{figure}

A detailed  numerical study  of  periodic clusters containing up 
to $L=12$ cells close to the  lines $FA$, $AC$, and $CD$ 
in Fig.~\ref{ph_diagram} suggests that 
for smaller values of  $J_{\perp}$
the ground state belongs to the sector $[1,0,\cdots ,1,0]$. 
In the special case  $K=0$,  our numerical results reproduce 
the phase diagram  of the frustrated SDC,\cite{takano}  
where the so-called tetramer-dimer phase (denoted as TD1 in 
Fig.~\ref{ph_diagram}) appears in the interval 
$0.909<J_{\perp}< 2$. This doubly degenerate 
singlet state  may be roughly represented as a product of   
single-diamond $S_2$ states on every second diamond, as depicted in 
Fig.~\ref{td_state}.  The  short-range correlations 
shown in Fig.~\ref{td_state} imply that the simple product state is a 
good variational wave function over the entire  region occupied by the  
TD1 phase, excluding a narrow region near the line AE. The product state 
is  an exact ground state  at  $(K,J_{\perp})=(0,2)$. An extrapolation of the 
exact-diagonalization (ED) data suggests 
that the phase boundary between the TD1 and  SF phases 
lies extremely close to the  $AC$ line separating the 
single-diamond  states $S_1$ and $S_2$. On the other hand, the phase boundary
between the TD1 and FM1 phases clearly deviates from the single-diamond 
boundary AF. The deviation from the line AF is related to weak but finite
interactions between neighboring $S_2$ diamonds appearing in the second-order 
perturbation theory in the parameter $J+K/4$.

On increasing the  cyclic exchange parameter $K$  at fixed
$J_{\perp}$ one finds  another tetramer-dimer type state (TD2). 
The picture  of the spin-spin correlations, Fig.~\ref{td_state}, suggests that 
every  second diamond is approximately  in the  $T_1$ state. 
Nevertheless, the numerical analysis shows relatively strong antiferromagnetic 
correlations between neighboring $T_1$ diamonds, as opposed to the TD1 state
where the $S_2$ diamonds are weakly correlated. Clearly, both  tetramer-dimer
phases are gapped and doubly-degenerated. 
According to the general rules,\cite{sachdev} one may expect a discontinuous 
quantum phase transition between the quantum gapped phases TD1 and TD2. 
Numerically, the transition point is indistinguishable from the
exact single-diamond   phase  boundary $CE$ ($K=4/5$) separating  the  
single-diamond states 
$S_2$ and $T_1$.  
\subsection{Phases in the sector 
$[1,1,\cdots ,1]$}
A numerical inspection of the short-range correlators
$\langle {\bf s_1}_n\cdot {\bf s_2}_n\rangle$ in finite periodic chains
implies that for moderate 
values of the parameter $J_{\perp}$ the established ground
states belong to the sector $[1,1,\cdots ,1]$. Thus, in the low-energy
region the diamond model, Eqs.~(\ref{h}) and (\ref{h2}), is equivalent
to  the following   mixed-spin
 Heisenberg model ($|{\bf s}_n|\equiv s_1=1$, $|\bs{\sigma}_n|\equiv s_2
=1/2$)  with   three-spin exchange interactions 
\beq\label{h3}
{\cal H}_1&=&\sum_{n=1}^L\left[
{\bf s}_n\cdot\left( \bs{\sigma}_n+\bs{\sigma}_{n+1}\right)
\! -\! J^{'}\bs{\sigma}_n\cdot
\bs{\sigma}_{n+1}\right.\nonumber \\
&+&\left. \frac{K}{2}\{ {\bf s}_n\cdot\bs{\sigma}_n\, ,\, 
{\bf s}_n\cdot\bs{\sigma}_{n+1}\}\right] \, ,
\eeq
where $J^{'}=3K/4-J$.
To the best of our knowledge, generic  mixed-spin quantum Heisenberg models 
with multiple-spin exchange interactions have not been discussed in the  
literature, although these  interactions may play an important
role in some recently synthesized mixed-spin magnetic
materials.\cite{kostyuchenko,note}
Here we  restrict ourselves to  
a general overview of the spin  phases in the specific case 
$(s_1,s_2)=(1,1/2)$, and  to ferromagnetic
exchange interactions between the $\bs{\sigma}_n$ spins  ($J^{'}>0$). 
The first two terms in  Eq.~(\ref{h3}) define a standard  mixed-spin
Heisenberg model containing additional non-frustrated  $J^{'}$ exchange bonds.
The Lieb-Mattis theorem\cite{lieb2} predicts a ferrimagnetic ground state 
for this bipartite model which coincides with the  classical ferrimagnetic 
two-sublattice  N\'eel state (the FM2 phase in Fig.~\ref{ph_diagram}). 
To study the role of the competing 
three-spin interactions, we use a  qualitative  spin-wave  analysis 
supplemented by  numerical ED  calculations  for finite periodic chains.
\begin{figure}[hbt]
\samepage
\begin{center}
\includegraphics[width=4cm]{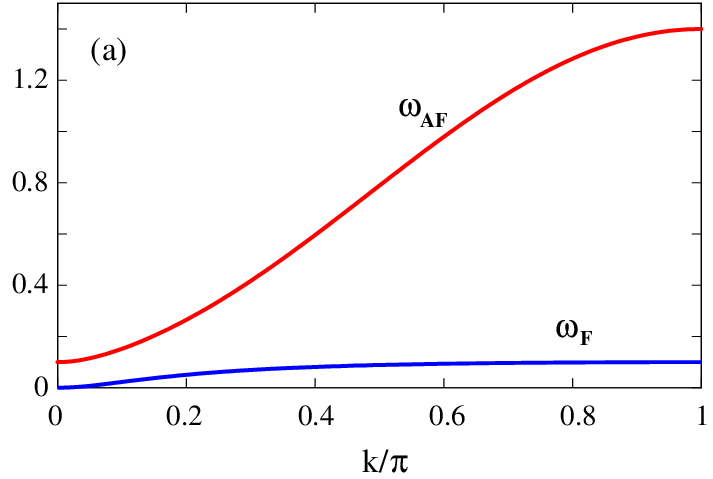}
\includegraphics[width=4cm]{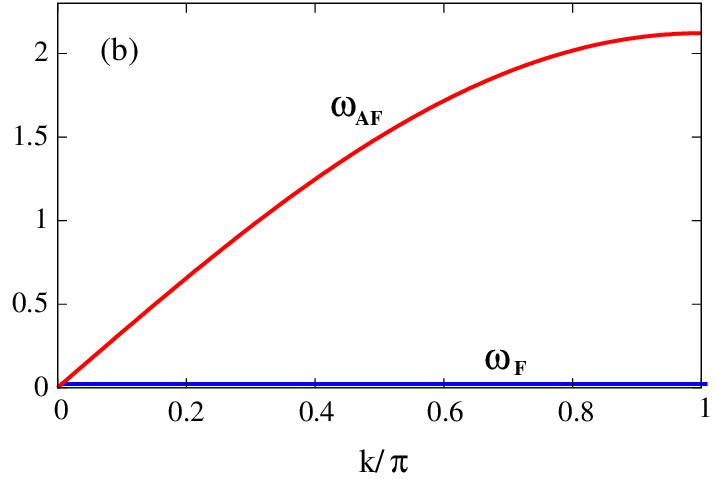}
\caption{\label{spectrum} Spin-wave spectrum of 
the $(s_1,s_2)=(1,1/2)$ chain  with three-spin exchange interactions, 
Eq.~(\ref{h3}),
in the ferrimagnetic (a) and the canted (b) ground states at $\kappa =0.8$ 
and $1.5$,
respectively. $J=J_{\perp}=0$, $\kappa \equiv K/2$.
}
\end{center}
\end{figure}                                      
\begin{figure}[hbt]
\samepage
\begin{center}
\includegraphics[angle=270, width=7cm]{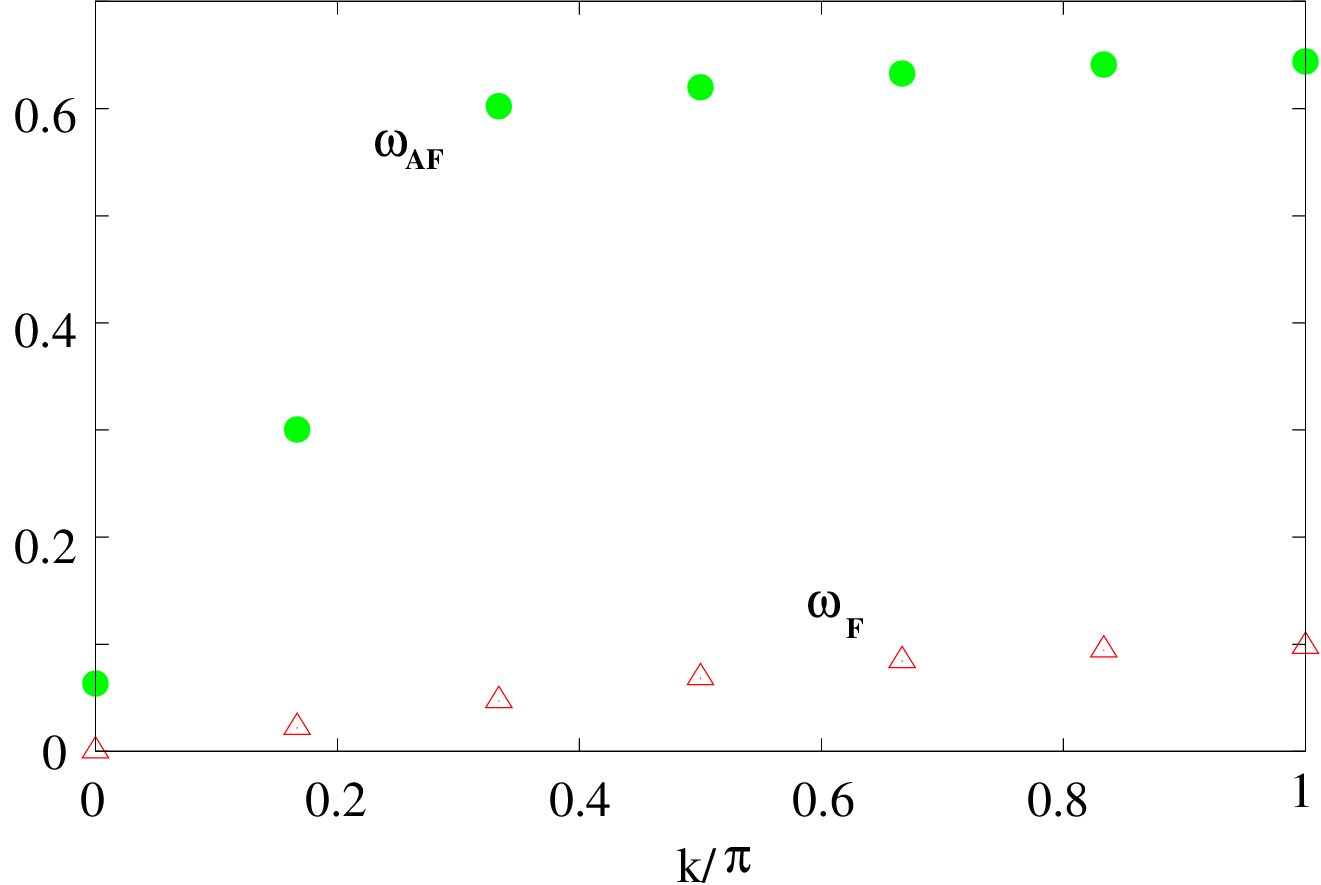}
\caption{\label{spectrum_ed} Lowest-energy excited states in a periodic 
chain with $L=12$ diamonds  in the sectors with magnetic moments $M_0-1$ 
(triangles) and $M_0+1$ (circles). $M_0=L(S_1-S_2)=6$ is the  magnetic 
moment of the ferrimagnetic ground state. $K=1$, $J=J_{\perp}=0$.
}
\end{center}
\end{figure}                                      

Let us  start with a discussion of the semi-classical limit of the model
(\ref{h3}).
Since the three-spin interaction contains an 
additional overall factor of $s_1s_2$,
it is convenient to redefine the coupling constant $K$ and measure the
strength of these interactions in terms of $\kappa = s_1s_2K$. 
The classical N\'eel configuration defined by ${\bf s}_n=s_1(0,0,1)$ and
$\bs{\sigma}_n=s_2(0,0,-1)$ survives up to $\kappa =1$. For
$\kappa >1$, the ferromagnetic arrangement of the smaller $s_2$ spins remains
unchanged whereas the orientation of the $s_1$ spins deviates from the z axis: 
${\bf s}_n=s_1(\sin \theta\cos \phi_n , \sin \theta \sin \phi_n , \cos 
\theta )$. Here  $\cos\theta =1/ \kappa$ and the azimuthal angle $\phi_n$ 
takes arbitrary values from  the interval $0\leq \phi_n<2\pi$ ($n=1,\ldots ,
L)$.  
As a rule, quantum fluctuations favor the planar spin  
configurations ($\phi_n =0,\pi$; $n=1,2,\ldots,L$),
so that one may expect a reduction of the degeneracy to $2^L$.  
Such high degeneracy of the ground state is typical for a number 
of spin models on corner-sharing  lattices. As seen in Fig.~\ref{spectrum}b,
in a spin-wave approximation the degeneracy produces a full line of
zero modes $\omega_{F}=0$ in the Brillouin zone.  
Note that the presence of the gapless antiferromagnetic mode
$\omega_{AF}\propto k$ ($k\ll 1$) for $\kappa > 1$  is related 
to the   finite transverse magnetization  of the classical canted state. 
The explicit expression for $\omega_{AF}$ reads
\be
\omega_{AF}=2J^{'}s_2\left| \sin \left(\frac{k}{2}\right) \right|
\sqrt{\sin^2 \left(\frac{k}{2}\right) +\frac{s_1}{s_2}\alpha\cos k}\, ,
\ee
where  $\alpha = (\kappa^2-1)/(2\kappa J^{'})$.
It is clear  that the antiferromagnetic  mode $\omega_{AF}$ 
is stabilized by the ferromagnetic couplings between the 
$\bs{\sigma}_n$ spins.  
The above picture of low-lying excitations in the ferrimagnetic phase
close to the phase transition point is confirmed by  the numerical results
at $K=1$,  Fig.~\ref{spectrum_ed}. The numerical estimate for 
the phase transition point  is  $K_c\approx 1.2$.
\begin{figure}[hbt]
\samepage
\begin{center}
\includegraphics[width=3.5cm]{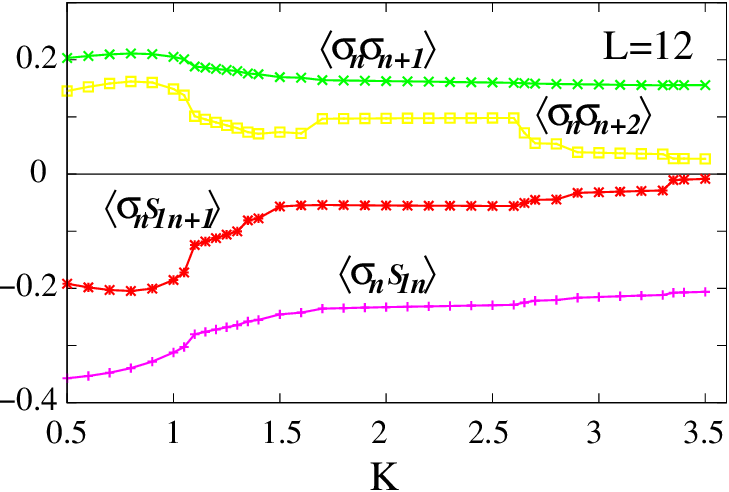}
\includegraphics[width=3.5cm]{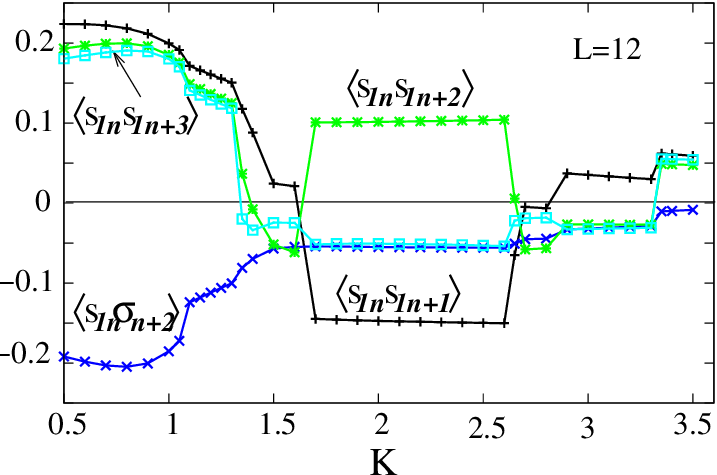}
\caption{\label{corr} Short-range spin-spin correlations vs. $K$
for periodic chains with $L=12$ cells ($J=J_{\perp}=0$).
}
\end{center}
\end{figure}                                      

\begin{figure}[hbt]
\samepage
\begin{center}
\includegraphics[width=7cm]{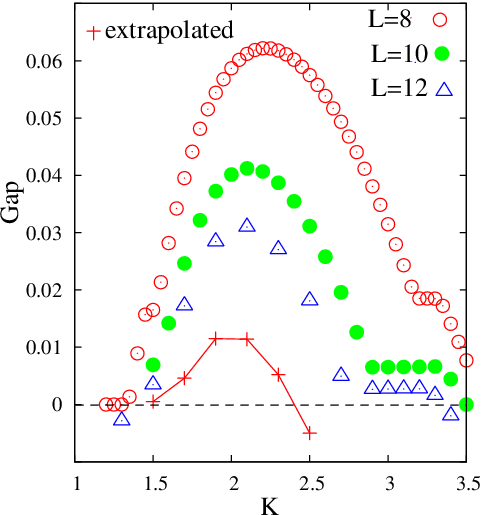}
\vspace{-.4cm}
\caption{\label{gap} The excitation gap in the DM phase
as a function of the parameter $K$.
}
\vspace{-.7cm}
\end{center}
\end{figure}

Additional information about the phase diagram  of Eq.~(\ref{h3}) may be 
obtained from the behavior of the short-range correlations, Fig.~\ref{corr}, 
and the singlet-triplet  excitation gap, Fig.~\ref{gap}, with the 
parameter $K$.  For larger $K>1.2$, the  ED results indicate at 
least two additional phases denoted by DM and FM3 in Fig.~\ref{ph_diagram}: 
DM is a non-magnetic gapped singlet state stabilized approximately 
in the region  
$1.5\leq K\leq 2.3$ (see  Fig.~\ref{gap}), and FM3 is a magnetic phase 
similar to the  FM2 phase.  Since the unit cell consists of three
spin-1/2 variables, the Lieb-Schultz-Mattis theorem\cite{lieb} suggests 
that the spin gap must be accompanied with at least doubly-degenerate ground 
states. 

\begin{figure}[hbt]
\samepage
\begin{center}
\includegraphics[width=7.5cm]{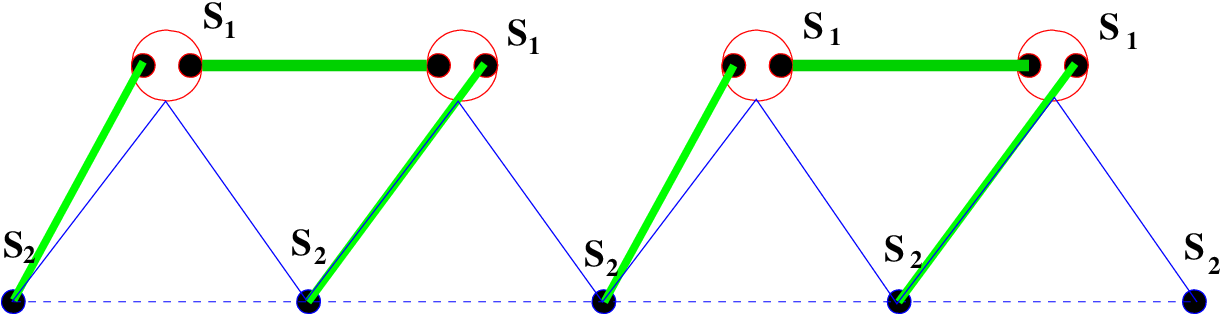}
\caption{\label{dimerization} Caricature of the dimerized singlet phase
DM as obtained from an extrapolation of the numerical ED results 
for periodic chains with $L=8,10$, and $12$ . The open circles (thick lines)
denote the  symmetric (antisymmetric) eigenstates of two
spin-1/2 operators. $K=2$, $J_{\perp}=J=0$, 
$(s_1,s_2)=(1,1/2)$.
}
\end{center}
\end{figure}

Further information about the DM phase can be extracted from
the dimerization operator $D_n={\bf S}_n\cdot{\bf S}_{n+1}-
{\bf S}_n\cdot{\bf S}_{n-1}$
where ${\bf S}_n$ is the spin operator at site $n$. The lattice 
sites $n-1$ and $n+1$ are supposed to be symmetric
under the reflection from the central site $n$. It is
convenient to use the  symmetric (antisymmetric) combinations
$|s,a\rangle =(|0\rangle\pm|1\rangle )/\sqrt{2}$, where $|0\rangle$ 
is the translationally invariant singlet ground state of the finite 
periodic chain  
and $|1\rangle$ is the singlet excited state which is  almost degenerate 
with $|0\rangle$. The states $|s,a\rangle$ are not translationally invariant 
and may be  expected to produce finite values of the dimer order-parameter 
$\langle s,a|D_n|s,a\rangle$ in a dimerized  system.\cite{chandra}
 The caricature of the DM state  presented in Fig.~\ref{dimerization} 
is obtained from the extrapolation 
of the ED results  for $\langle s,a|D_n|s,a\rangle$ 
($L=8,10$, and $12$). A pronounced enhancement of the dimer order-parameter 
 with $L$ is indicated only for two types of bonds in the dimer model
(the thick lines in  Fig.~\ref{dimerization}). Clearly, the suggested
ground state exhibits a four-fold degeneracy.   
%
%

Finally,  the numerical results point towards the existence of another
partially polarized magnetic phase  in the narrow interval $1.2<K <1.5$
between the ferrimagnetic (FM2) and the dimerized (DM) phases
(the  hatched area in Fig.~\ref{ph_diagram}). Recently, similar exotic magnetic 
states have been predicted in a number of one-dimensional spin systems with
magnetic frustrations.\cite{ivanov2} Typically, discussed spin states
exhibit   a partially polarized magnetization in the $z$ direction 
($M< M_0=s_1-s_2$), a  quasi-long-range transverse magnetic order, and a
gapless linear  mode related to the destroyed classical
canted state.   Unfortunately, the methods used in the 
present study do not suggest a clear statement indicating the existence of such an exotic phase in the discussed  
system.     
    
\section{Summary}
In conclusion, we have examined the impact of the cyclic four-spin 
exchange interactions on the ground-state phase diagram of the symmetric 
spin-1/2 diamond chain. Using the local symmetries of the   model, 
the spin phases were classified  by the set of good quantum numbers 
$s_n=0,1$ ($n=1,2,\dots,L$) related to the composite spins  
${\bf s}_n={\bf s_1}_n+{\bf s_2}_n$. The presented study  demonstrates 
a rich phase diagram in the parameter space $(K,J_{\perp})$.  
Apart from  the standard magnetic and paramagnetic phases, the system 
exhibits two different  tetramer-dimer phases 
in the sector $[1,0,1,0,\cdots,1,0]$ as well as an exotic  
four-fold-degenerate dimerized ground state in the 
sector $[1,1,\cdots,1]$. 

\begin{acknowledgments}
 This research was supported by Deutsche
Forschungsgemeinschaft (Grant 436 BUL 17/9/06)
and the Bulgarian Science Foundation (Grants
F1414 and D002-264/18.12.08). Part of the work has been  done in 
Max-Planck-Institut f\"ur Physik komplexer Systeme,
Dresden.
\end{acknowledgments} 


\end{document}